# Giant twist-angle dependence of thermal conductivity in bilayer graphene originating from strong interlayer coupling

H. F. Feng, B. Liu, and Zhi-Xin Guo*

State Key Laboratory for Mechanical Behavior of Materials, Center for Spintronics and Quantum System, School of Materials Science and Engineering, Xi'an Jiaotong University, Xi'an, Shaanxi, 710049, China.

*zxguo08@xjtu.edu.cn

**Abstract**

Recently, the twist-angle effect on 2D van der Walls (vdW) materials, such as bilayer graphene, has attracted great attention. Many novel electronic, magnetic and even optical properties induced by such effect have been discovered. However, the twist-angle effect on phononic property is not so remarkable. By investigating the thermal conductivity of twist bilayer graphene (TBG), here we reveal that the trivial twist-angle effect on phononic property observed in previous studies is owing to the non-localization nature of phonons. This characteristic makes phonons hardly trapped by the weak interlayer potentials induced by the twist-angle dependent Moiré pattern. We propose that the twist-angle effect can be effectively enhanced by increasing the interface coupling. In use of a sandwich structure composed of h-BN and TBG, we demonstrate that the thermal conductivity of TBG can be either significantly increased or dramatically decreased, under the synergistic modulation of interlayer coupling strength and twist angle. Particularly, the twist-angle effect can lead to a nontrivial reduction of thermal conductivity up to 78% when a strong interlayer coupling is applied. The reduction is several times larger than that observed in the freestanding TBG where the reduction is attributed to the twist-angle dependent phonon scatterings induced by the edge phonons. The underlying mechanism for the giant twist-angle dependence of thermal conductivity is further revealed on the basis of phonon transport theory. Our findings provide a platform for achieving efficient twist-angle modulation on phonon transport property of vdW materials.

During the past decade, twisted two-dimensional (2D) materials have attracted extensive attentions in both academic research and engineering applications, because of their unique and adjustable physical and chemical properties [1-9]. Compared with strong intralayer covalent bonding interaction, the weak van der Walls (vdW) interlayer interaction in 2D materials makes it possible to stack and assemble the vdW homogeneous/heterogeneous systems layer by layer, which provides powerful means for designing and manufacturing 2D stacked materials with intriguing properties [10,11].

In stacked vdW materials, the interlayer twist angle (denoted as $\psi$) is of a special concern in recent years, where the twist bilayer graphene (TBG) has been the most widely investigated materials in both experimental and theoretical studies [12-15]. Especially, a flat electronic band with vanishing Fermi velocity appears at Dirac points when $\psi$ becomes particularly small (so called magic angle) [16]. As a result, the relative strong coupling between interlayer electrons around the moiré point gives rise to certain filling states of the flat band and thus a series of novel properties on electrons [17,18], magnets [19-22], photons [23-25], etc. For instance, metal-insulator transition and superconducting states appear with $\psi=1.1°$ and $1.05°$, respectively [18,26,27]. Emergent superstable ferromagnetism was discovered with $\psi=1.2°$ [19]. In addition, the moiré patterns of TBG becomes a natural plasmon photonic crystal for propagating nano-light with $\psi=0.06°$ [28].

On the other hand, the twist-angle effect of phononic property is not so attractive [29]. Although extensive theoretical and experimental investigations have been carried out under various conditions, the twist-angle effect only leads to a maximum reduction of thermal conductivity (K) in TBG by about 30% [30-33]. Moreover, K does not obviously depend on the twist angle, but has a V-shaped relationship with the commensurate lattice constant ($\lambda$) with the lowest K appearing around 1 nm [31]. As far as we know, why the twist-angle effect of phononic property is much inferior to the electronic correlated ones is still to be clarified. In addition, from both fundamental research and practical application points of view, it is in great need to find an efficient way that can effectively enhance the twist-angle effect of phononic property.

By investigating K of TBG based on the molecular dynamics simulations and phonon transport theory, here we reveal the intrinsic weak twist-angle effect of phononic property. We find the trivial twist-angle effect on K is owing to the nature of non-localization of phonons, which makes phonons hardly trapped by the weak vdW interlayer potentials induced by the moiré pattern. We also find that the twist angle-effect on phonons can be effectively enhanced by increasing the interlayer coupling, which is demonstrated in a sandwich structure composed of h-BN and TBG.

To verify the twist angle effect on K, we first consider a TBG with 90 nm in length (that is, 30nm ×10nm in x-y plane with heat flux in x direction) and adopt periodic boundary condition in the width (y) direction (denoted as 2D TBG). Here we mainly consider the TBG with $\psi= 21.79°$, $13.17°$, $9.43°$ and $5.09°$, respectively [Figs. 1(a)-1(d)], which present distinguishable moiré patterns [34]. To explore the magic angle effect on K, we additionally consider the TBG with $\psi=1.08°$. As indicated in Fig. 1(e), the non-equilibrium molecular dynamics (NEMD) method is used to obtain the thermal conductivity of TBG [See Supplementary Material for details]. Note that the 10 nm width is large enough to get a converged K in the periodic boundary condition according to our simulations. Fig. 2 (a) shows twist angle dependent K of TBG with different temperatures. It is found that K hardly changes with twist angle at room temperature (300 K). Although the twist-angle effect becomes a little more significant at low temperature region (50 K), the difference of K is still very small, with the maximum variation being less than 9.6% (from 3540 W/mK to 3200 W/mK). Such weak twist-angle effect on K, which is negligible in comparison with the experimental observations (20%-30% variation of K at room temperature), is unexpected.

It is noted that the periodic boundary condition in the width direction is adopted in our calculations, where the phonon scatterings induced by the edge phonons in width direction is eliminated. This may be the reason for the much weaker twist-angle effect on K in the 2D TBG. Note that in the experimental case, the TBG always has edge phonons due to its finite width. To verify such speculation, we further calculate K of TBG nanoribbons (30nm ×10nm in x-y plane), where the shrink-wrapped boundary condition is applied in the width direction. As shown in Fig. 2(b), K first decreases and then increases with the twist angle increasing, reaching the minimum value (200W/mK) at 13.17°, 37% lower than that at 0° (317W/mK). This result is obviously different from that obtained in 2D TBG, but agrees well with the experimental observations and other theoretical calculations.

Fig. 2(c) additionally shows the temperature distributions in the length direction of the simulated TBGs. A common feature is that TBG nanoribbons have obviously larger temperature gradient than 2D TBGs. Moreover, the temperature gradient in 2D TBGs is nearly independent of the twist angle, in opposite to that in TBG nanoribbons. In Figs. 2(d)-2(f) we additionally show the heat flux distributions in 2D TBG and TBG nanoribbons. Different from the 2D TBG with a uniform large heat flux distribution [Fig. 2(d)], there are remarkable non-uniform and twist-angle dependent heat flux distributions in TBG nanoribbons. As shown in Figs. 2(e) and 2(f), the heat flux becomes particularly small around the edges of nanoribbon, showing the nature of strong boundary scatterings induced by the edge phonons which leads to the reduction of heat flux. Notably, the heat flux of TBG nanoribbon with $\psi$=21.79° is much smaller than that of 0°, which is distinct even in the center region, showing that the edge phonons can lead to significant reduction of heat flux in the whole TBG region. The underlying mechanism can be attributed to the strong scatterings between edge phonons and body phonons, the extent of which dependents on the specific edge structures determined by the twist angles. Therefore, the 20%-30% variation of K reported in previous studies is in fact due to the edge phonons, which have localized characteristics.

The above results on the other hand show that the vdW interlayer potential induced by moiré pattern is too weak to trap phonons, which results in little change of K. This is because phonons are collectively excited quasi particles in nature, which are hardly localized in a defect-free system. Hence, such weak twist-angle effect is not expected for other particles (such as electrons, magnets, photons), the singularity of which are strongly correlated to the localized electrons nearby the moiré points [16,35]. This feature also explains the absence of flat band in the phonon structures with a magic angle (1.05º) [36] that had been widely observed in the electronic structures [18].

Accordingly, exotic twist-angle effect on K of TBG can be expected when the interlayer interaction is effectively enhanced. It is known that an easy way to enhance interlayer interaction is reducing the interlayer distance ($d_{int}$) by pressurizing. So we construct a h-BN/TBG/h-BN sandwich structure, where $d_{int}$ can be effectively reduced by decreasing the distance between the h-BN layers under and above TGB, respectively. As shown in Fig. 3(a) and 3(b), two sandwich structures are considered: (1) the h-BN layers synchronously rotate with their adjacent graphene layers (denoted as Cor structure), so that the rotation angle between h-BN layer and its adjacent graphene is always zero as shown in Fig. 3(c). (2) The h-BN layers keep unrotated during the rotation of graphene layers (denoted as Uncor structure). In this case, additional moiré patterns between h-BN and its adjacent graphene would be formed due to the nonzero twist angle [Fig. 3(d)]. Moreover, in order to clearly exhibit the interlayer interaction effect on K of TBG, the vibration of h-BN layers is frozen during the simulation. A 30nm×10nm sandwich structure is considered in the simulation, where the periodic boundary condition is applied in the width direction to eliminate the influence of edge phonons.

Fig. 4(a) shows the calculated twist-angle dependence of K of TBG in both Cor and Uncor structures at 300 K. In order to make a direct contradistinction, K of freestanding TBG is also shown in Fig. 4(a), which exhibits slight variation with the twist angle due to the perturbation of simulated cell length. A common feature is that the twist-angle dependence of K in both Cor and Uncor structures is very similar to that in the freestanding TBG when the interlayer interaction is weak ($d_{int}$ = 3.4Å, 3.1Å). This result agrees with our exception that the weak interlayer interaction cannot lead to significant twist-angle effect on K, even with the additional interface interaction effect from h-BN. Nevertheless, an interesting phenomenon is observed, i.e., K of TBG is obviously increased in the sandwich structures. To eliminate the influence from perturbation of simulated cell length, we further calculate the change-ratio of K, which is defined as

$$\Delta K = \frac{K_1 - K_0}{K_0} \times 100\% \quad (1)$$

where $K_0$ and $K_1$ are the twist-angle dependent thermal conductivity of TBG in freestanding and sandwich structures, respectively. As shown in Fig. 4(b), on one hand, $\Delta K$ has little twist-angle dependence in the sandwich structures. On the other hand, $\Delta K$ obviously increases with the reduction of $d_{int}$ from 3.4Å to 3.1Å, where a sizeable increment of about 10% can be obtained with $d_{int}$= 3.1Å in both Cor and Uncor structures. The unusual increase of K can be attributed to the interlayer interaction from h-BN, which changes the phonon structure and thus leads to the reduction of phonon-phonon scatterings (Supplementary Material).

When the interlayer distance is further reduced to $d_{int}$=2.8 Å (interlayer interaction is 4.4 times larger than that of $d_{int}$=3.1 Å), the variation of K becomes dramatic. As shown in Fig. 4(a), the TBG with $\psi$=0º (AB stacking) is significantly larger than that with $\psi$>0º in both Cor and Uncor structures. Note that the interface coupling makes K of TBG in Cor and Uncor structures has contrary behavior: it is increased in Cor structure but significantly decreased in the Uncor structure, owing to the different interface interactions from h-BN. The maximum K appears in Cor structure (852 W/mK) with $\psi$=0º, which is about 27% larger than that of freestanding TBG (672 W/mK). This result shows that interface coupling can be an efficient way to increase of K of a bilayer system, which has great potential applications in the heat management in chips. On the other hand, as shown in Fig. 4(b), $\Delta K$ remarkably decreases by about 23% with $\psi$ increasing from 0º to 5.09º, and then slightly increases by about 4% as $\psi$ increases from 5.09º to 21.79º. This result verifies that the increase of interlayer interaction to certain extent can effectively enhance the twist angle effect on phonon transport properties, due to the significant modulation of phonon structure and thus phonon-phonon scatterings induced by the moiré pattern potential (Supplementary Material).

More exotic variation of K can be obtained when the interlayer interaction is further increased. As shown in Fig. 4(a), with $d_{int}$ decreased to 2.5Å (interlayer interaction becomes 17.2 times larger than that of $d_{int}$=3.1 Å), in both Cor and Uncor structures K is extraordinarily decreased. The maximum K in Cro and Uncor structures is 447 W/mK and 301 W/mK, respectively, which is 33% and 55% smaller than the freestanding TBG. Note that in the maximum K in Cor structure appears with $\psi$=21.79º. This is different from the weaker interlayer interaction cases as well as that in the Uncor structure, where the the maximum K exists with $\psi$=0º. Meanwhile, the minimum K in both Cor (139 W/mK) and Uncor (167 W/mK) structures appears at 5.09º, with $\Delta K$=78% and 73%, respectively. This result shows that giant reduction of K in vdW materials can be easily realized under the synergistic effect of strong interlayer interaction and nonzero twist angle, which may have potential applications in thermal insulations and thermoelectrics.

It is noticed that the distinguished twist angle dependent of K can be induced by strong interlayer interactions in both Cor and Uncor structures. In order to explore the maximum variation of K induced by the twist angle effect, we further calculate the maximum difference of $\Delta K$ which is defined as

$$\gamma = \frac{K_{max} - K_{min}}{K_{max}} \times 100\% \qquad (2)$$

where $K_{max}$ and $K_{min}$ are the maximum and minimum K of TBG with various the twist angles but the same $d_{int}$, respectively. It is found that $\gamma$ in the Cor and Uncor structures reaches up to 69% and 45%, respectively, significantly larger than that originated from edge effects in TBG nanoribbions (20-30%). This simulation result also agrees well with our proposed mechanism, that is, the strong interlayer interaction can remarkably enhance the effect of moiré pattern potential on phonons. Note that the twist angle effect in Cor structure is much more profound than that in Uncor structure, especially in the nonzero twist angle region. This is because there is an additional twist angle (or moiré pattern) effect in Uncor structure, originated from the relative rotation between graphene and its nearby h-BN layers (as indicated in Fig. 3). The "double" moiré patterns make K in TBG easily saturated to a certain value and thus robust to nonzero twist angles under the strong interlayer interactions.

Then we come to reveal the unusually large reduction of K at small twist angle (5.09º) under the strong interlayer interaction in Cor structure, which results in the outstanding twist angle effect. We first show in Fig. 5(a) the temperature distributions of 5.09º and 21.79º Cor structures, which have the smallest and largest K, respectively. It is found that there is large temperature jump in the 5.09º structure, while it is absent in the 21.79º structure. A temperature jump usually means the existence of thermal interface that hinders the phonon transport.

To unveil the reason for such strange phenomenon, we further extract the atomic structures during the NEMD simulation at 300K. As shown in Fig. 5(b), there is indeed an additionally small length moiré pattern appearing in the middle of the simulated cell in the 5.09º structure (here we call it "moiré junction"), which transfers the simulated cell to be homojunction structure. Note that such homojunction structure is purely induced by the relative rotation between two TBG layers, because there is no structural distortion in the in-plane direction of each graphene layer (See Fig. S3). Moreover, the moiré junction is formed only in the TBG with small and nonzero twist angles under the strong interlayer interaction. As shown in Fig. 5(c), the moiré junction is absent in the large angle (21.79°) structure, so that the temperature distribution is very smooth in the thermal conduction region. This is because the structures with small twist angle is usually unstable to that with large twist angles. As a result, the phase transition more easily happens in the small twist-angle structures under the effect of external stress.

Fig. S4 additionally shows the Cor structures of TBG with $\psi=5.09°$ and 21.79° when $d_{int}$ = 2.8Å. In comparison with the structures when $d_{int}$ = 2.8Å, obvious phase transition appears in the 5.09° structure when $d_{int}$ = 2.5Å which results in moiré junction. Whereas, the phase transition does not happen when $d_{int}$ = 2.8Å. Fig. 5(d) additionally shows the phonon density of states (PDOS) in the moiré junction, compared with that in the remaining regions as marked in Fig. 5(b). It is seen that significant mismatch exists between the two PDOS in the low energy region (below 3.5THz), indicating the distinct phonon structures in the two TBG regions. Consequently, the low-energy phonons will receive additional scatterings when propagating across the moiré junction, which leads to the thermal interface and thus reduction of K [37].

Before conclusion, it should be pointed out that the h-BN/TBG/h-BN sandwich structures considered in the simulation are based on a simplified model where the vibration of h-BN layers is artificially frozen. Nevertheless, the obtained physics is still valid in the real systems. For example, we have additionally considered the multilayer h-BN/TBG/multilayer h-BN structure where the h-BN layers nearby TBG can freely vibrate [Figs. S5(a) and 5(b)]. As shown in Figs. S5(c) and S5(d), the obtained twist angel and $d_{int}$ dependence of K is very similar to the simplified model.

In summary, we reveal the intrinsic mechanism for the twist-angle effect on phonon transport properties. Based on the NEMD simulations for both TBG nanoribbons and 2D TBG, we clarify that the trivial twist-angle effect on phononic property of 2D TBG is owing to the nature of non-localization of phonons. It makes phonons hardly trapped by the moiré-pattern resulted interlayer potentials. The slight twist-angle dependent thermal conductivity of TBG observed in experiments in fact originates from the phonon scatterings induced by the edge phonons which have obvious twist-angle dependence. We additionally propose that the twist-angle effect on phonons can be effectively enhanced by increasing the interface coupling, which can be realized in the h-BN/TBG/h-BN sandwich structure. The NEMD simulations show that the thermal conductivity of TBG can be either obviously increased by 27% or remarkably decreased by several times under the synergistic effect of interlayer coupling strength and twist angle. Notably, the twist-angle effect can lead to a reduction of thermal conductivity up to 69% under strong interface interaction, several times larger than that induced by the edge phonons in the freestanding TBG nanoribbons. The underlying mechanism for these interesting results is further explained based on the phonon transport theory. Finally, we would like to remark that the giant twist-angle dependence of thermal conductivity induced by strong interface interactions should be generally applied to other twisted vdW materials, such as bilayer transition metal oxygens or dichalcogenides and buckled graphene-like structures.

We thank Prof. Xin-Gao Gong for valuable discussions. This work was supported by the Natural Science Foundation of China (Grant No. 12074301). We gratefully acknowledge the computational resources provided by the HPCC platform of Xi'an Jiaotong University.

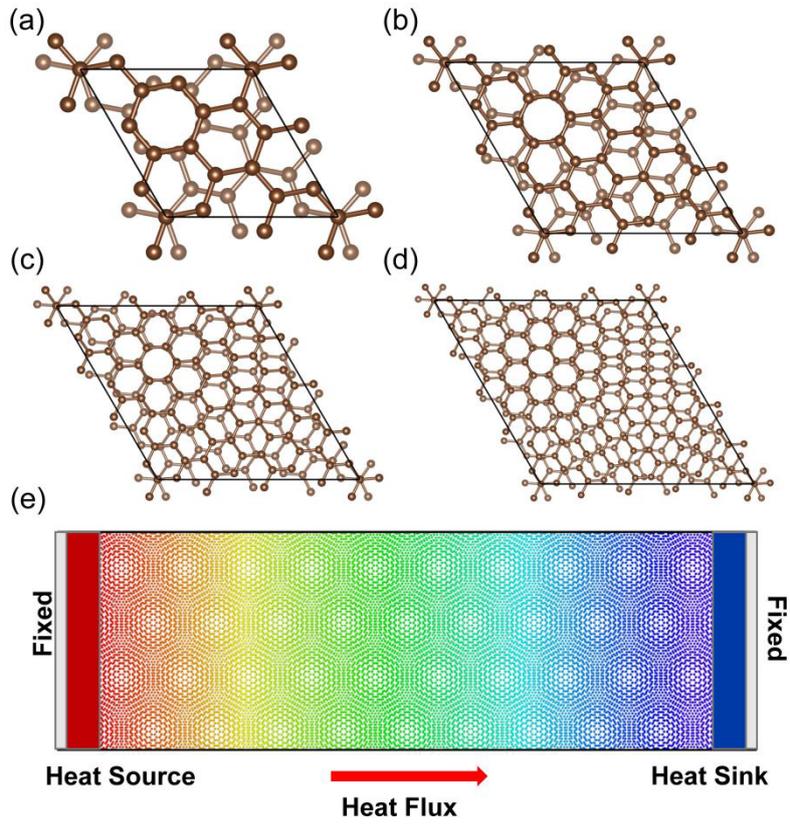

Fig. 1. Top views of the primitive cells of TBGs with various twist angels ($\psi$). (a) $\psi$=21.79°; (b) $\psi$=13.17°; (c) $\psi$=9.43°; (d) $\psi$=5.09° TBG. (e) NEMD model with simplified scales of in-plane conduction.

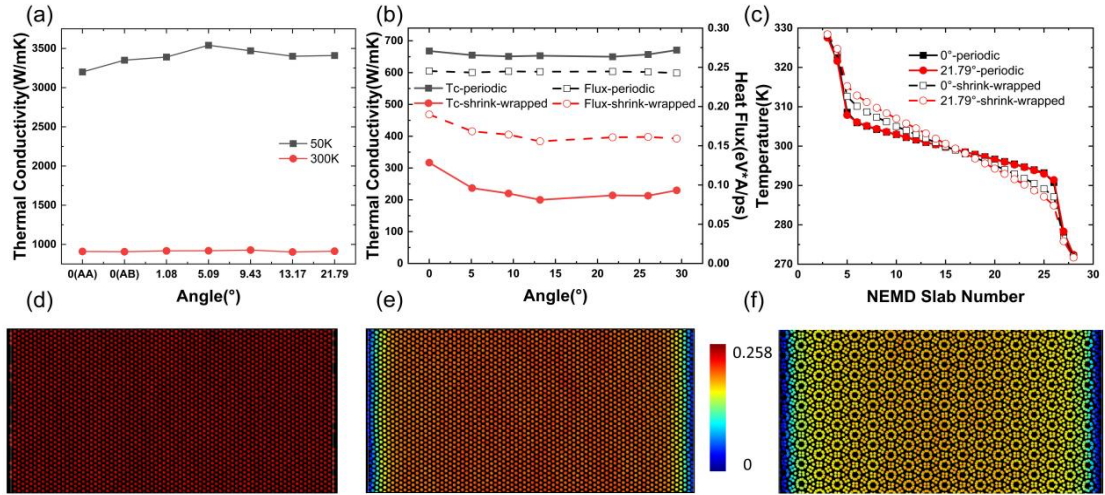

Fig. 2. (a) Relationship between thermal conductivity of TBG and angles at different temperature (90 nm×10nm). (b) Thermal conductivity and heat flux of TBG as a function of twist angle (30 nm×10nm). (c) Temperature profile as a function of NEMD slab index for 2D TBG and TBG nanoribbons. Spatial distribution of heat flux accumulation of (d) $\psi=0°$ for 2D TBG, (e) $\psi=0°$ for TBG nanoribbon, (f) $\psi=21.79°$ for TBG nanoribbon.

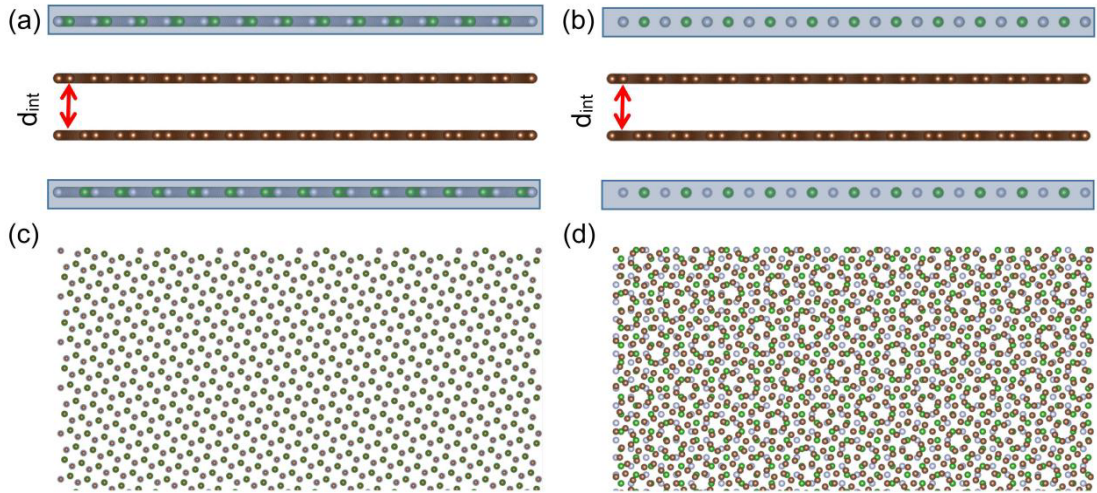

Fig.3. Schematic structures of TBG sandwiched between two fixed h-BN layers of both Cor and Uncor structures. (a) top view of Cor structure, (b) top view of Uncor structure, (c) side view of Cor structure, (d) side view of Uncor structure. $d_{int}$ represents the interlayer distance between two graphene layers. The Green, saddle brown and gray balls represent the B, C and N atoms, respectively. The fixed h-BN layers are indicated by the semitransparent rectangle.

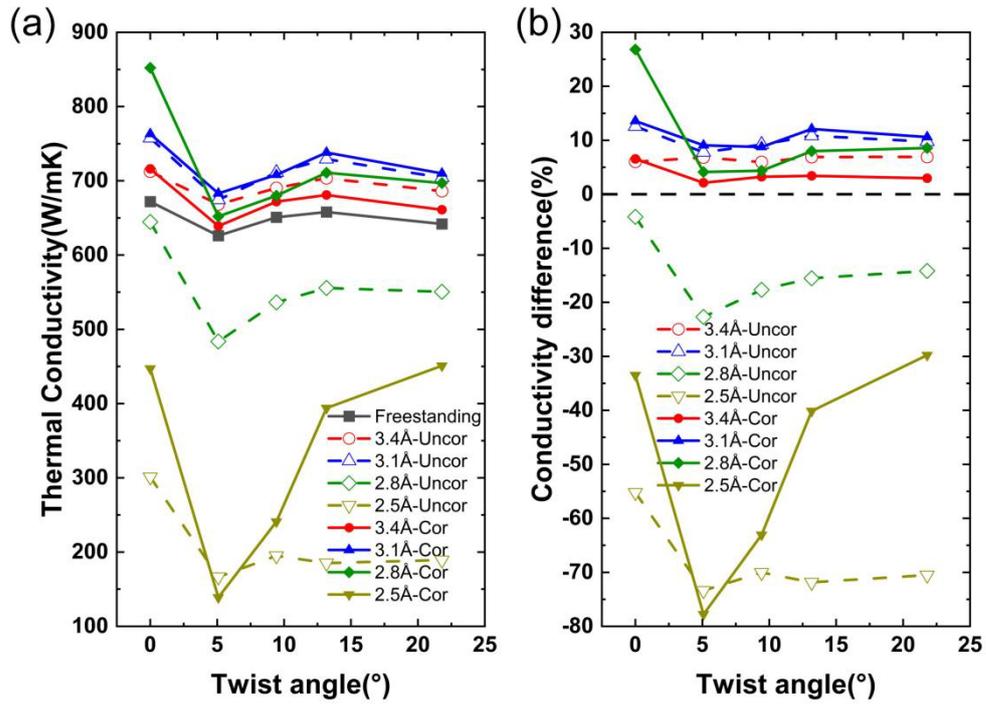

Fig. 4. (a) Thermal conductivity and (b) Conductivity difference of various configurations as a function of twist angle.

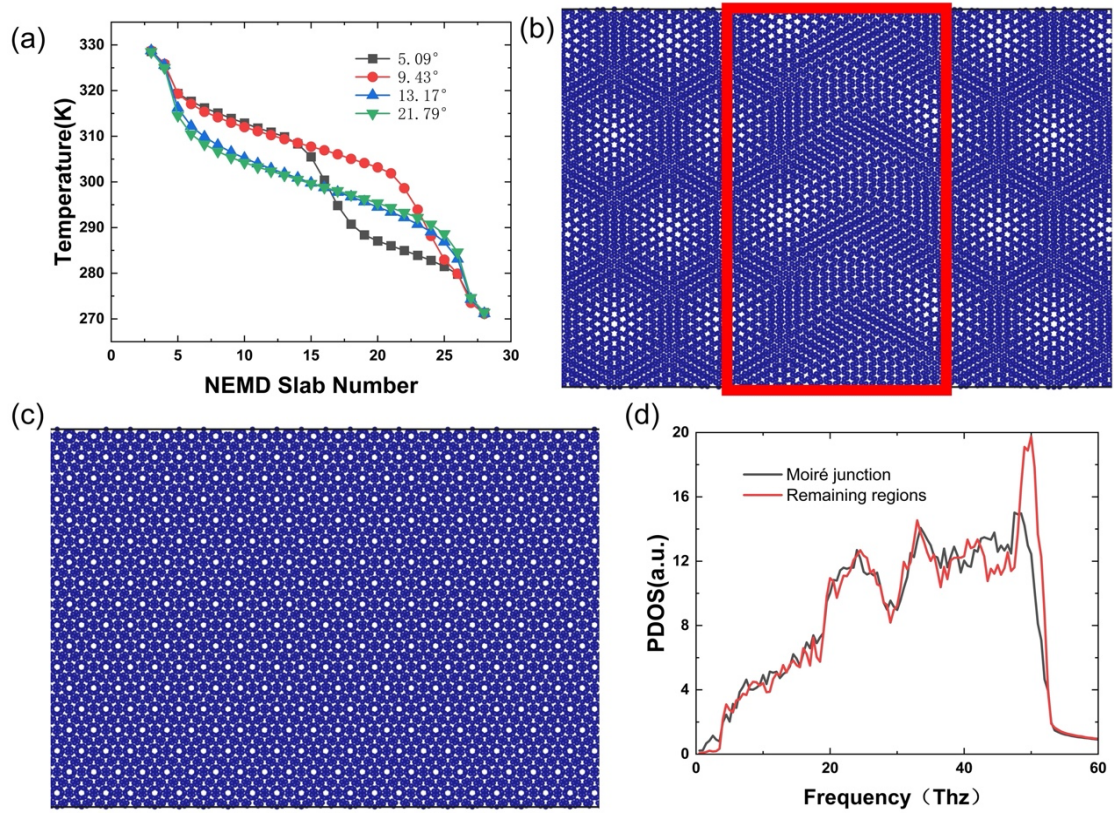

Fig. 5. (a) Temperature profile as a function of NEMD slab index for Cor structures with various twist angles at $d_{int}$ = 2.5Å. (b) Schematic Cor structure with $\psi$=5.09° for $d_{int}$ = 2.5Å. (c) Schematic Cor structure with $\psi$=21.79° for $d_{int}$ = 2.5Å. (d) PDOS of Moiré junction (indicated by the red quadrangle) and the normal region with $\psi$=5.09° for $d_{int}$ = 2.5Å.